\documentclass[final]{iopart}

\expandafter\let\csname equation*\endcsname\relax
\expandafter\let\csname endequation*\endcsname\relax

\usepackage{amsmath}
\usepackage{hyperref}
\usepackage{iopams}
\usepackage[dvips]{graphicx}
\usepackage{color}

\begin{document}

\title{Broadband and ultra-broadband polarization rotators with adiabatic modular design}
\author{Emiliya Dimova$^1$, Andon Rangelov$^2$$^,$$^3$, and Elica Kyoseva$^3$}

\begin{abstract}
We experimentally demonstrate a broadband and an ultra-broadband spectral bandwidth polarization rotators. Both polarization rotators have modular design, that is, they are comprised of an array of half-wave plates rotated to a given angle. We show that the broadband and ultra-broadband performance of the polarization rotators is due to the adiabatic nature of the light polarization evolution. In this paper we experimentally investigate the performance of broadband and ultra-broadband polarization rotators comprising of ten multi-order half-wave plates or ten commercial achromatic half-wave plates, respectively. The half-wave plates in the arrays are rotated gradually with respect to each other starting from an initial alignment between the fast polarization axis of the first one and the incoming linearly polarized light, to the desired polarization rotation angle.
\end{abstract}



\address{$^1$ Institute of Solid State Physics, Bulgarian Academy of Sciences, 72
Tsarigradsko chauss\'{e}e, 1784 Sofia, Bulgaria}

\address{$^2$ Department of Physics, Sofia University, James Bourchier 5 blvd., 1164
Sofia, Bulgaria}

\address{$^3$ Engineering Product Development, Singapore University of Technology and
Design, 8 Somapah Road, 487372 Singapore}

\ead{rangelov@phys.uni-sofia.bg}




\section{Introduction}


The analogy between the Poincar\'{e} space, which is used for the description of light polarization, and the Bloch space, which is used for the description of quantum-state dynamics of two-level systems, is well-known \cite{Kubo1980,Kubo1981,Kubo1983,Kuratsuji1998,Kuratsuji2007,Ardavan,Rangelov2010}. This has allowed a transfer of concepts from the field of quantum optics \cite{Allen,Shore} and nuclear magnetic resonance \cite{Levitt&Freeman,Levitt,Freeman} to the field of polarization optics \cite{Azzam,Goldstein}. For instance, the concept of composite pulses \cite{Levitt&Freeman,Levitt,Freeman} was successfully used to create broadband half- and quarter-wave plates \cite{Ivanov,Peters,Dimova2014} as well as tunable polarization rotators \cite{Rangelov2015,Dimova2015}. Another example is the application of adiabatic techniques developed for quantum systems in polarization optics \cite{Zapasskii,Rangelov2011,Berent,Shore2015} to design various robust and achromatic polarization retarders and optical isolators.

In this paper we report an experimental demonstration of a broadband and an ultra-broadband polarization rotators, designed by an analogy to piecewise adiabatic passage technique \cite{Shapiro2007,Shapiro2008,Zhdanovich}. The broadband and ultra-broadband polarization rotators are assembled according to the recent theoretical work of Shore et al. \cite{Shore2015} as a sequence of ten multi-order half-wave plates and ten commercial achromatic half-wave plates, respectively. For both designs, the half-wave plates are rotated gradually from an alignment between the fast-polarization axis with the initial linearly polarized light to an alignment of the fast-polarization axis with the desired final linear polarization orientation.


\section{Theory}


We first consider the propagation of a plane electromagnetic wave along the
z-axis through a lossless medium. In this case polarization evolution is
given with the torque equation for the Stokes vector \cite%
{Kuratsuji1998,Kuratsuji2007,Sala,Seto}:
\begin{equation}
\frac{\,\text{d}}{\,\text{d}z}\mathbf{S}(z)=\mathbf{\Omega }(z)\times
\mathbf{S}(z),  \label{Stokes equation}
\end{equation}%
where $z$ is the distance along the propagation direction, $\mathbf{S}%
(z)=[S_{1}(z),S_{2}(z),S_{3}(z)]$ is the Stokes vector, and the driving
torque $\mathbf{\Omega }(z)=[\Omega _{1}(z),\Omega _{2}(z),\Omega _{3}(z)]$
is the birefringence vector of the medium. When the medium does not possess
optical activity and is uniaxial with the slow and fast axes in the $xy$
plane, then the components of the birefringence vector $\mathbf{\Omega }(z)$
are given explicitly as
\begin{eqnarray}
\Omega _{1}(z) &=&\Omega _{0}\cos \left( 2\varphi \right) , \\
\Omega _{2}(z) &=&\Omega _{0}\sin \left( 2\varphi \right) , \\
\Omega _{3}(z) &=&0, \\
\Omega _{0} &=&\frac{2\pi }{\lambda }\left( n_{e}-n_{o}\right) .
\end{eqnarray}%
Here $n_{e}$ and $n_{o}$ are the refractive indices along the fast and slow
axes, $\lambda $ is the light wavelength, $\Omega _{0}$ is the rotary power,
and $\varphi $ is the angle of rotation between the fast (slow) axis and the
$x$ ($y$) Cartesian axis.

Now if the Stokes vector $\mathbf{S}$\ is initially parallel to the birefringence vector $\mathbf{\Omega }$, such that $\mathbf{\Omega }\times\mathbf{S=0}$, and we change slowly (adiabatically) the angle $\varphi $, then the Stokes vector will also evolve adiabatically with it. For instance, if linearly polarized light is initially in the horizontal plane, i.e., $\mathbf{S}\left( z_{i}\right) =\left( 1,0,0\right) $, and $\varphi \left( z_{i}\right) =0$, which corresponds to aligning the $x$ and $y$ axes of the Cartesian coordinate system to coincide with the fast and slow optical axes, then the birefringence vector $\mathbf{\Omega }$ is initially aligned with the Stokes vector. Following an adiabatic evolution, the final angle of linear polarization will be $\varphi \left( z_{f}\right) $. This adiabatic approach has the advantages of being robust with respect to large intervals of values for $\Omega_{0}$  and is therefore, broadband.

\begin{figure}[tbh]
\centerline{\includegraphics[width=0.8\columnwidth]{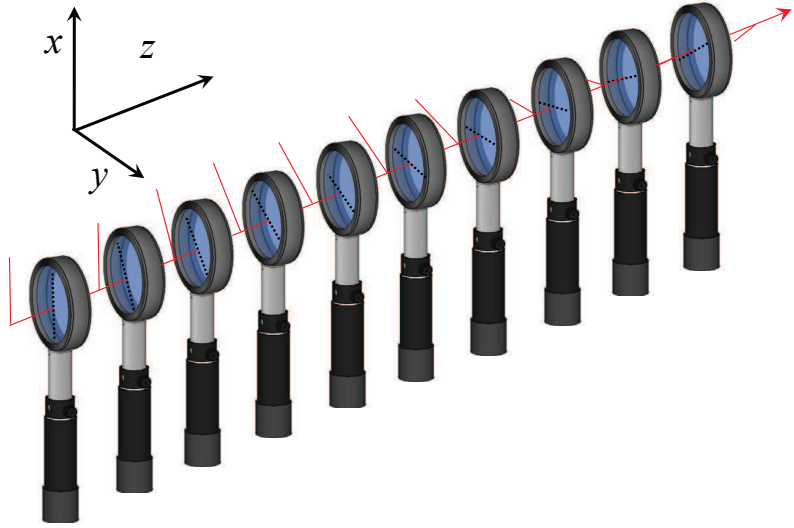}}
\caption{(Color online) Schematic illustration of a discrete adiabatic light
evolution, which consists of a series of birefringent modules, separated by
free-space propagation. The fast polarization axes of the wave plates are
represented by dashed lines, while the solid red lines represent the
orientation of linear polarization.}
\label{fig:fig1}
\end{figure}

In this work we propose an alternative scheme for the adiabatic evolution for the Stokes vector by means of discrete modular design of the polarization rotators \cite{Shore2015,Shapiro2007,Shapiro2008,Zhdanovich}. That is, we consider that the rotation of the angle $\varphi $ takes place in a sequence of discrete steps, rather than continuously. Experimentally, we realize this by a sequence of $N=10$ birefringent crystals, each rotated at an angle $%
\varphi _{m}$ with respect to the chosen Cartesian coordinate system
(illustrated schematically in Figure \ref{fig:fig1}). For such a scheme the overall
adiabatic evolution is achieved if the individual rotation angles change gradually
and if there are sufficient number of steps ($\varphi _{m}=\frac{\varphi
}{N}\ll \varphi $). It is necessary that each of the birefringent crystals
drives not more than a "Rabi\textquotedblright\ cycle, in other words, that each
module is not more than a half-wave plate.

\section{Experiment}

\subsection{Optical setup}
\label{Optical setup}
We performed several experiments to verify our theoretical predictions for the performance of both broadband and ultra-broadband adiabatic polarization rotators.
The linear polarization of white light beam was rotated at 45$^{0}$, 60$^{0}$, 75$^{}$ and 90$^{0}$ degrees passing through a set of ten half-wave plates. 
\begin{figure}[tbh]
\centerline{\includegraphics[width=0.9\columnwidth]{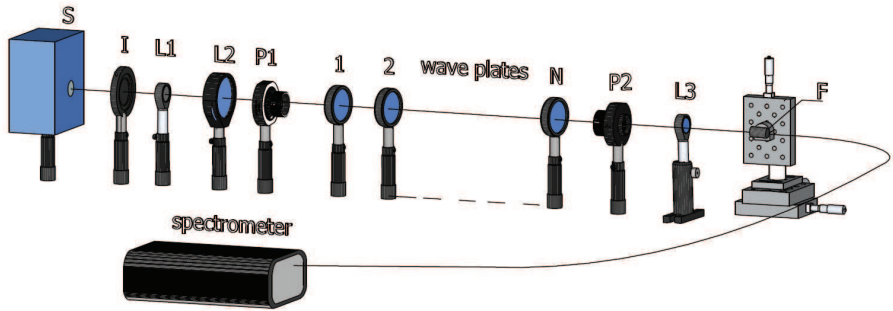}}
\caption{(Color online) Experimental setup. The source $S$, irises $I$, lens
$L_{1}$, lens $L_{2}$ and polarizer $P_{1}$ form a collimated beam of white
polarized light. Analyzer $P_{2}$ and lens $L_{3}$ focus the beam of output
light onto the entrance $F$ of an optical fibre connected to a spectrometer.
The stack of ten wave plates rotates the linear polarization of the light in the broadband and ultra-broadband spectra. }
\label{fig:fig2}
\end{figure}
A 10 W Halogen-Bellaphot (Osram) lamp with DC power supply was applied as a light source with continuous spectrum in the range from 450 nm to 1100 nm (Figure \ref{fig:fig2}). A sequence of an iris, two lenses and a polarizer were used for production of collimated light beam. The iris $I$ was placed in the focus of plano-convex lens $L_{1}$ ($f$=35mm) imitating a point light source. The received beam was additionally collimated by a second lens $L_{2}$ with $f$=150 mm. Polarizer $P_{1}$ (Glan-Tayler, 210-1100 nm,  borrowed from a Lambda-950 spectrometer) linearly polarized the white light in the horizontal plane.

We used two types of half-wave plates for our experiments. In the first experiment a set of ordinary multi-order quarter-wave plates (WPMQ10M-780, Thorlabs) which perform as half-wave plates at $\lambda$=763 nm were used while in the second experiment achromatic half-wave plates (WRM053-mica, 700-1100 nm, aperture 20 mm) were exercised. Each wave plate (aperture of $1^{\prime\prime}$) was assembled onto a separate RSP1 rotation mount which realizes a $360^{0}$ rotation. The scale marked at $2^{0}$ increments allows for precise, repeatable positioning and fine angular adjustment.

We analyzed the characteristics of the transmitted light through the ten wave plates by a polarizer $P_{2}$ (the same type as $P_{1}$) and a grating monochromator (Model AvaSpec-3648 Fiber Optic Spectrometer with controlling software AvaSoft 7.5). We used a plano-convex lens $L_{3}$ ($f$=20 mm) and a two-axis micro-positioner to focus the light beam onto the optical fibre entrance $F$ of the monochromator.
\begin{figure}[tbh]
\centerline{\includegraphics[width=0.8\columnwidth]{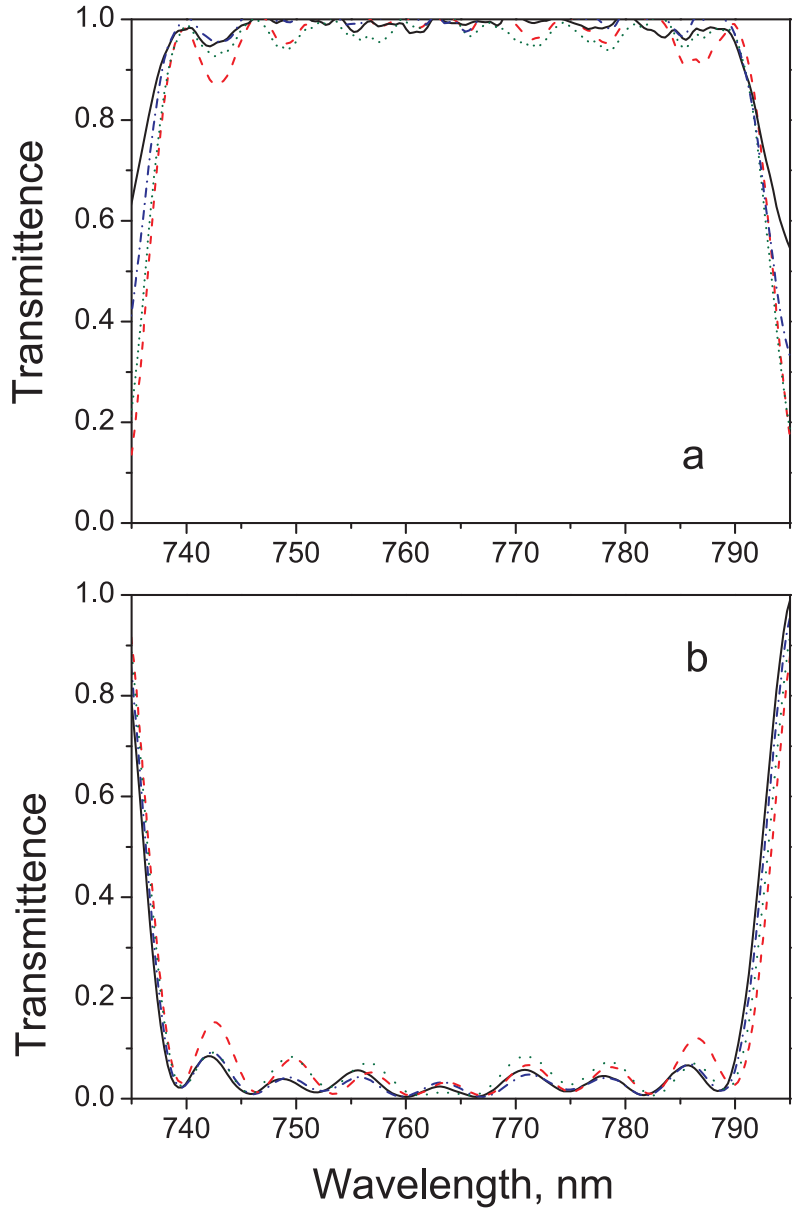}}
\caption{(Color online) Broadband linear polarization rotator consisting of ten multi-order half-wave plates. The curves corresponded to different rotation angles $\alpha$ (a) and $\alpha+90^{0}$ (b) as follow: red dashed line $90^{0}$, green short dash line $75^{0}$, blue dashed dotted line $60^{0}$ and black solid line $45^{0}$.}
\label{fig:fig3}
\end{figure}

\subsection{Measurement procedure}
 Our aim was to demonstrate broad and ultra-broad spectral bandwidth linear polarization rotators composed by two sets of ten identical wave plates. Each set was assembled by multi-order wave plates or by achromatic wave plates, respectively, as described above in Subsection \ref{Optical setup}. The wave plates were slightly tilted  to reduce unwanted reflections \cite{Peters}. For the analysis of the experimental data we used a single beam spectrometer. To account for noise and losses due to light transmission, reflection and absorption in different media, we measured the light and dark spectra for all experiments. Furthermore, we used the dark spectrum, which is measured with the light path completely blocked, for correction of hardware offsets. The reference spectrum is usually taken with the light source on and a blank sample instead of the sample of interest measured.  In our case, however, we measured the transmission spectrum of the already assembled polarization rotator, but  the axes of the polarizer $P_{1}$, the analyzer $P_{2}$ and the fast axis of each single wave plates were all set parallel. We used the measured light spectrum as a reference for the subsequent measurements.

A linearly polarized light beam passed by the sequences of $N$ wave plates. Each of them was adjusted at the estimated angle by the formula:
$\varphi_{m}=\left(m-1\right)\frac{\alpha}{\left(N-1\right)}$,
where $m$ is the sequential number of the wave plate and $N=10$ in the present experiment. The rotation effect was investigated for each of the arbitrary chosen angles $\alpha=\left\{45^{0}, 60^{0}, 75^{0}, 90^{0}\right\}$. The data were taken with analyzer $P_{2}$ revolved at angle $\alpha$ and $\alpha+90^{0}$.
\subsection{Experimental results}
\begin{figure}[tbh]
\centerline{\includegraphics[width=0.8\columnwidth]{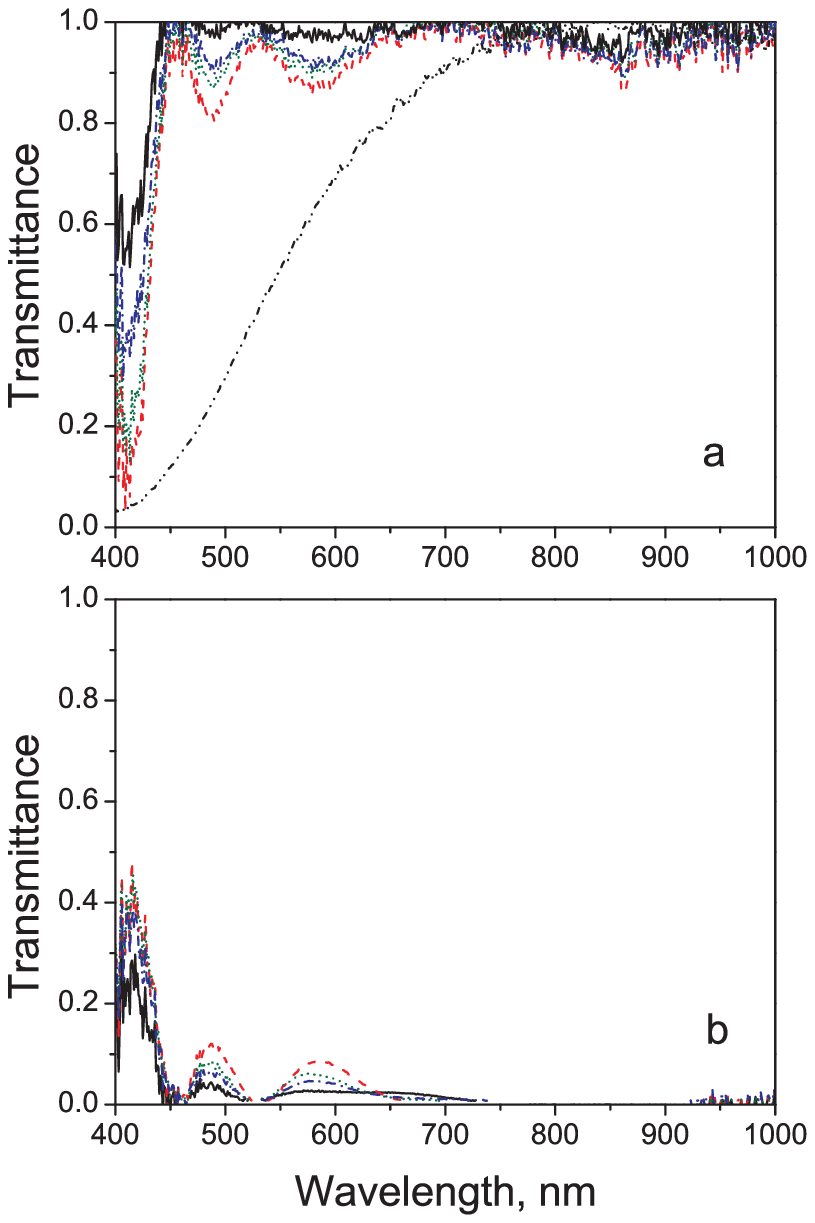}}
\caption{(Color online) Ultra-broadband linear polarization rotator built by ten achromatic half-wave plates. The curves corresponded to different rotation angle $\alpha$ (a) and $\alpha+90^{0}$ (b) as follow: red dashed line $90^{0}$, green short dash line $75^{0}$, blue dashed dotted line $60^{0}$ and black solid line $45^{0}$. The black dashed dot dot curve shows the spectrum of single achromatic wave plate.}
\label{fig:fig4}
\end{figure}
We experimentally tested and proved development of a broad spectral bandwidth polarization rotator settled by ten multi-order half-wave plates. The transmittance spectra were taken for four arbitrary chosen angles $\alpha$. The  obtained results are presented on the Figure~\ref{fig:fig3} a. The experimental curves show flat maximum in wide spectral diapason what is kept at different angles $\alpha$.  The experimental curves at angles $\alpha+90^{0}$ of the analyzer $P_{2}$ were also presented on Figure~\ref{fig:fig3} b. The blow out the ray at $\alpha+90^{0}$ proves that the polarization of the light is orthogonal to the one at angle $\alpha$.

Even broader effect was reached by using ten achromatic half-wave plates. We created ulta-broad bandwidth linear polarization rotator following the above explained algorithm. We used the same rotation angles $\alpha$. Each of the wave plate in the sequences was the same type and with axis rotated at the angle $\varphi_{m}$. The experimental results are shown on the Figure~\ref{fig:fig4} a. Again for proving that the developed set of wave plates was working as a linear polarization rotator, the transmittance spectra at angles $\alpha+90^{0}$ are presented (see Figure~\ref{fig:fig4} b). The shown experimental curves not exceed 1000 nm because of the lack of the signal after this spectral region.


\section{Conclusion}

In this paper we demonstrated experimental application of piecewise adiabatic passage, concept introduced by Shapiro et al. in quantum optics
\cite{Shapiro2007,Shapiro2008,Zhdanovich}, for an optical broadband and ultra-broadband polarization rotators. The broadband and ultra-broadband
polarization rotators are modular designed as stacks of multi order half-wave plates or commercial achromatic half-wave plates, which are rotated smoothly from alignment between the fast-polarization axes with initial linear polarized light to the alignment of the fast-polarization axes with desired linearly polarization orientation. The experimental results show visible broadening effects which are kept the same at different angles of the chosen sequence of wave plates.

\section*{Acknowledgements}

E. D acknowledges financial support by Bulgarian National Science Fund Grant: DRila 1/04. A.R and E.K acknowledge financial support by SUTD start-up Grant No. SRG-EPD-2012-029, SUTD-MIT International Design Centre (IDC) Grant No. IDG31300102.



\begin{thebibliography}{99}

\bibitem{Kubo1980} Kubo H and Nagata R 1980 \emph{Opt. Commun.} \textbf{34} 306

\bibitem{Kubo1981} Kubo H and Nagata R 1981 \emph{J. Opt. Soc. Am.} \textbf{71} 327

\bibitem{Kubo1983} Kubo H and Nagata R 1983 \emph{J. Opt. Soc. Am.} \textbf{73} 1719

\bibitem{Kuratsuji1998} Kuratsuji H and Kakigi S 1998 \emph{Phys. Rev. Lett.} \textbf{80} 1888

\bibitem{Kuratsuji2007} Kuratsuji H, Botet R and Seto R 2007 \emph{Prog. Theor. Phys.} \textbf{117} 195

\bibitem{Ardavan} Ardavan A 2007 \emph{New J. Phys.} \textbf{9} 24

\bibitem{Rangelov2010} Rangelov A A, Gaubatz U and Vitanov N V 2010 \emph{Opt. Commun.} \textbf{283} 3891

\bibitem{Allen} Allen L and Eberly J H 1987 \emph{Optical Resonance and Two-Level Atoms} (Dover, New York)

\bibitem{Shore} Shore B W 1990 \emph{The Theory of Coherent Atomic Excitation} (John Wiley \& Sons, New York)

\bibitem{Levitt&Freeman} Levitt M H and Freeman R 1979 \emph{J. Magn. Reson.} \textbf{33} 473

\bibitem{Levitt} Levitt M H 1986 \emph{Prog. Nucl. Magn. Reson. Spectrosc.} \textbf{18} 61

\bibitem{Freeman} Freeman R 1997 \emph{Spin Choreography} (Spektrum, Oxford)

\bibitem{Azzam} Azzam M A and Bashara N M 1977 \emph{Ellipsometry and Polarized Light} (North Holland, Amsterdam)

\bibitem{Goldstein} Goldstein D and Collett E 2003 \emph{Polarized Light} (Marcel Dekker, New York)

\bibitem{Ivanov} Ivanov S S, Rangelov A A, Vitanov N V, Peters T and Halfmann T 2012 \emph{J. Opt. Soc. Am. A} \textbf{29} 265

\bibitem{Peters} Peters T, Ivanov S S, Englisch D, Rangelov A A, Vitanov N V and Halfmann T 2012 \emph{Appl. Opt.} \textbf{51} 7466

\bibitem{Dimova2014} Dimova E, Ivanov S S, Popkirov G and Vitanov N V 2014 \emph{J. Opt. Soc. Am. A} \textbf{31} 952

\bibitem{Rangelov2015} Rangelov A A and Kyoseva E 2015 \emph{Opt. Commun.} \textbf{338} 574

\bibitem{Dimova2015} Dimova E, Rangelov A and Kyoseva E \emph{to be published} (arXiv:1502.00747)

\bibitem{Zapasskii} Zapasskii V S and Kozlov G G 1999 \emph{Phys. Usp.} \textbf{42} 817

\bibitem{Rangelov2011} Rangelov A A 2011 \emph{Opt. Lett.} \textbf{36} 2716

\bibitem{Berent} Berent M, Rangelov A A and Vitanov N V 2013 \emph{J. Opt.} \textbf{15} 085401

\bibitem{Shore2015} Shore B W, Rangelov A A, Vitanov N V and Bergmann K \emph{to be published}

\bibitem{Shapiro2007} Shapiro E A, Milner V, Menzel-Jones C and Shapiro M 2007 \emph{Phys. Rev. Lett.} \textbf{99} 033002

\bibitem{Shapiro2008} Shapiro E A, Pe'er A, Ye J, Shapiro M 2008 \emph{Phys. Rev. Lett.} \textbf{101} 023601

\bibitem{Zhdanovich} Zhdanovich S, Shapiro E A, Shapiro M, Hepburn J W and Milner V 2008 \emph{Phys. Rev. Lett.} \textbf{100} 103004

\bibitem{Sala} Sala K L 1984 \emph{Phys. Rev. A} \textbf{29} 1944

\bibitem{Seto} Seto R, Kuratsuji H and Botet R 2005 \emph{Europhys. Lett.} \textbf{71} 751

\end{thebibliography}
\end{document}